\documentstyle[12pt,epsf]{article}
\newcommand{\bflambda}{\mbox{\boldmath $\lambda$}}
\newcommand{\bfsigma}{\mbox{\boldmath $\sigma$}}
\newcommand{\bftau}{\mbox{\boldmath $\tau$}}
\newcommand{\bfphi}{\mbox{\boldmath $\phi$}}
\newcommand{\bfa}{\mbox{\boldmath $a$}}
\newcommand{\bfk}{\mbox{\boldmath $k$}}
\newcommand{\bfp}{\mbox{\boldmath $p$}}
\newcommand{\bfr}{\mbox{\boldmath $r$}}

\begin{document}

\begin{center}

{\large\bf
Nucleon resonances in the constituent quark model
with chiral symmetry}

\vspace{0.5cm}
{\large H. Kowata, M. Arima}

{\it Department of Physics, Osaka City University, Osaka 558, Japan}

\vspace{0.5cm}
{\large K. Masutani}

{\it Faculty of Liberal Arts and Education, 
Yamanashi University, Kofu 400, Japan} 

\end{center}

\begin{abstract}
 The mass spectra of nucleon resonances with spin $\frac{1}{2}$, 
$\frac{3}{2}$, and $\frac{5}{2}$ are systematically 
studied in the constituent quark 
model with  
meson-quark coupling, which is inspired by the spontaneous breaking 
of chiral symmetry of QCD. 
 The meson-quark coupling gives rise not only to the 
one-meson-exchange potential between quarks but also to the self-energy 
of baryon resonances due to the existence of meson-baryon 
decay channels.
  The two contributions are consistently taken into account 
in the calculation.
  The gross properties of the nucleon resonance spectra are 
reproduced fairly well although the predicted mass of $N(1440)$ 
is too high. 
\end{abstract}

\begin{quote}
11.30.Rd, 12.39.Jh, 14.20.Gk
\end{quote}


\section{Introduction}
 Although the constituent quark models have been successful in reproducing
various static properties of baryons \cite{HK,Isgur-Karl},
there still remains several problems.
 One of them is the mass spectrum of the spin $\frac{1}{2}$ 
nucleon resonances. 
The conventional models cannot explain the fact that the ground state 
is remarkably light while the separations among the excited states are 
relatively narrow \cite{PDG}. Furthermore, the mass of $N(1440)$
is predicted to be too high. 

 Glozman {\it et al.} have recently examined baryon mass spectra 
by using the constituent quark model with  the one-meson-exchange 
potential (OMEP) as the residual interaction between 
quarks \cite{Glozman-Riska,GPP,GPVW},
whereas the conventional models contain the one-gluon-exchange 
potential (OGEP) instead \cite{RGG}. 
 The OMEP has been introduced on the basis of the spontaneous 
breaking of chiral symmetry (SB$\chi$S) in low-energy QCD. 
In direct consequence of SB$\chi$S,  there appear Nambu-Goldstone 
bosons (i.e., the flavor-octet pseudoscalar mesons such 
as $\pi$, $K$, and $\eta$) 
as well as the mass of light quarks \cite{MG,CL}.
 In addition to the flavor-octet mesons, 
the flavor-singlet $\eta '$ has been also
taken into account for the Nambu-Goldstone bosons.
 Glozman {\it et al.} have claimed that the model provides a unified
description of the ground states and the excitation spectra of baryon
resonances. They have also pointed out that the spin-flavor 
dependence of the OMEP is important to reproduce the mass spectra.
 
 On the other hand, there exists another type of mesonic effects.
The self-energy of baryons, which comes from the coupling 
between single baryon states and meson-baryon scattering states, 
have been studied by one of the authors (M.A.) and his 
collaborators \cite{ASY,HAM,HMA}.  
They have shown, for example, that the large $\eta N$ decay
width of $N(1535)$ as well as  the mass splitting between 
$\Lambda(1405)$ and $\Lambda(1520)$ can be explained by the 
effect of the self-energy. We note here that they have used the OGEP 
for the residual quark-quark interaction. 

 In this paper, we construct the model consisting of the constituent 
quarks and the pseudoscalar mesons with the pseudovector meson-quark 
coupling.  The quarks and mesons are treated as the elementary 
degrees of freedom on the basis of SB$\chi$S.
 Since the meson-quark coupling gives rise not only to the OMEP but also 
to the self-energy, the two mesonic effects can be treated on an 
equal footing. 
 The OMEP mainly stems from the exchange of the off-energy-shell mesons 
whose energies are nearly equal to zero, 
while the self-energy does from the on-energy-shell mesons. 
 The two contributions, however, cannot be strictly separated as is 
clearly seen in the relativistic formalism where they correspond 
to the same diagram \cite{Chin}.  Therefore, 
the problem of double counting should be resolved   
in order to treat them in a consistent manner.
 The purpose of this work is not to obtain a perfect fit to the 
observed baryon masses but to investigate the consistent model containing 
the two mesonic effects.  
  We calculate the mass spectra of nucleon resonances and 
mainly examine the dynamical effect on the mass spectrum 
of the self-energy, which is absent in the conventional calculations. 
 The self-energy provides the energy-dependent effect on resonance 
masses, while the OMEP does the static effect.  

The paper is organized as follows. In Section 2, the constituent
quark model with the meson-quark coupling is presented, and the OMEP
and the self-energy are derived with an emphasis on the
problem of double counting.  The method of calculation is explained also.  
The numerical results for nucleon resonances 
are shown in Section 3. 
The summary of the paper is given in Section 4.

\section{Model}
 The model Hamiltonian for the meson-quark system $H$ consists of the 
internal Hamiltonian of the baryon $H_0$, the kinetic energy 
of the baryon  $T_{B}$ (i.e., the center-of-mass (c.m.) motion of 
the three quarks), the total energy of the meson $\omega_{M}$, 
and the meson-quark coupling $H_{I}$, as follows:    
\begin{equation}
H = H_0 + T_{B} + \omega_{M} + H_{I}\ ,
\label{eqn:hamiltonian}
\end{equation}
with 
\begin{equation}
H_{0} = T_{0} + V_{\rm conf} + M_{0}\ ,
\label{eqn:hamiltonian0}
\end{equation}
where $T_{0}$ is the kinetic energy of the quarks without the 
c.m. motion, $V_{\rm {conf}}$ the confinement potential for the quarks, 
and $M_{0}$ the constant mass parameter of the baryon.
 Note that $T_{B}$ and $\omega_{M}$ have non-vanishing 
contributions only for meson-baryon scattering channels.

 The non-relativistic kinematics is used for the constituent quarks:
\begin{equation}
T_{0} + T_{B} = \sum_{i=1}^3\frac{\bfp_i^2}{2m} ,
\label{eqn:hamiltonian1}
\end{equation} 
where $m$ is the quark mass and 
$\bfp_i$ the momentum of the $i$th quark.
Here the mass difference between u- and d-quarks is 
neglected and the conventional value of the quark mass is used:
$m = 340\,\rm{MeV}$.
The above decomposition can be easily done by means of Jacobi
coordinates in the non-relativistic kinematics.  This property 
is quite favorable in the investigation of baryon mass 
spectra and  meson-baryon scattering. 

 For the confinement potential that ensures the three 
quarks are always confined in the baryon, the linear form is employed:
\begin{equation}
V_{\rm conf} = c\sum_{i<j}^3|\bfr_i-\bfr_j|\ ,
\label{eqn:hamiltonian2}
\end{equation} 
where $c$ is the strength parameter and 
$\bfr_i$ the coordinate of the $i$th quark.
This type of confinement is suggested by the lattice QCD calculations 
for heavy quark systems \cite{Creutz} as well as by the 
flux tube model \cite{Kogut}.
 Since light constituent quarks are considered in this work, $c$ should 
be treated as a phenomenological parameter. 
 The bare mass $M_{0}$ is a free parameter also.

 For the meson-quark coupling $H_{I}$, 
the non-relativistic form of the pseudovector coupling is employed:
\begin{eqnarray}
H_{I} 
&=& 
\sum^3_{i=1} 
  \left\{
    \frac{ig_{Mqq}}{2m}\rho(\bfk)
    \left[
      \frac{\omega_{M}}{2m}
      \left(\bfsigma_i\cdot\stackrel{\leftarrow}{\bfp_i}
      \bflambda_i\cdot\bfphi_i(\bfk)
      \right.
    \right.
   \right.
\nonumber\\
&&+
   \left.
     \left.
       \left. 
         \bflambda_i\cdot\bfphi_i(\bfk)\,
         \bfsigma_i\cdot\stackrel{\rightarrow}{\bfp_i}
       \right)
       -\bfsigma_i\cdot\bfk\,
       \bflambda_i\cdot\bfphi_i(\bfk)
     \right]
   \right.
\nonumber\\
&&+ 
   \left.
     \frac{ig_{\eta'{qq}}}{\sqrt{6}m}\rho(\bfk)
     \left[
       \frac{\omega_{\eta'}}{2m}
       \left(\bfsigma_i\cdot\stackrel{\leftarrow}{p}_i
          \phi^{\eta'}_i(\bfk)+\phi^{\eta'}_i(\bfk)\,
          \bfsigma_i\cdot\stackrel{\rightarrow}{p}_i
       \right)
     \right.
   \right.
\nonumber\\
&&-
   \left.
     \left.
       \bfsigma_i\cdot\bfk\,\phi^{\eta'}_i(\bfk)
     \right]
   \right\}+{\rm{h.c.}}\ ,
\label{eqn:pv_coupling}
\end{eqnarray}
with
\begin{equation}
\rho(\bfk) = \exp\left(-\frac{k^2}{8 a^2}\right)\ ,
\end{equation}
\begin{equation}
\bfphi_i(\bfk) =
\frac{1}{\sqrt{2\omega_{M}}}\bfa_i(\bfk)\,
e^{ik}\cdot\bfr_i\ ,
\end{equation}
\begin{equation}
\phi^{\eta'}_i(\bfk) =
\frac{1}{\sqrt{2\omega_{\eta'}}}\,a^{\eta'}_i(\bfk)\,
e^{i\bfk\cdot\bfr_i}\ ,
\end{equation}
where 
$g_{Mqq}$ and $g_{\eta ' qq}$ are the meson-quark 
coupling constants for the flavor-octet and flavor-singlet mesons,
respectively; 
$\bfsigma_i$, $\bflambda_i$, and 
$\stackrel{\rightarrow}{\bfp}_i$ ($\stackrel{\leftarrow}{\bfp}_i$) are 
the spin SU(2) operator, the flavor SU(3) operator, and 
the initial (final) momentum operator of the $i$th quark, respectively. 
 The flavor-octet meson field with the momentum 
$\bfk=\stackrel{\leftarrow}{\bfp}_i-\stackrel{\rightarrow}{\bfp}_i$ 
is represented by $\bfphi_i(\bfk)$ with the meson annihilation 
operators on the $i$th quark $\bfa_i(\bfk)$.
 The flavor-singlet meson field is similarly denoted 
by $\phi^{\eta'}_i(\bfk)$. 
 The phenomenological form factor $\rho(\bfk)$ is included to incorporate 
the finite size of the meson and the constituent quark.

 The pseudovector coupling is one of the lowest-order terms 
with respect to the derivative operator 
of the meson field in the effective theory based on chiral 
symmetry \cite{Campbell}.
 This type of coupling has the favorable properties to reproduce the 
low-energy $\pi N$ phase shift and the $\eta$ production cross section 
around the $\eta N$ threshold \cite{ASY,Bethe-Hoffmann}.
 Note also that the pseudovector coupling brings about the same OMEP
between quarks as the pseudoscalar coupling.

 As mentioned above, meson-baryon continuum states have to be 
considered due to the meson-quark coupling $H_{I}$,
in addition to the single baryon described as a three-quark 
bound state.
 Multi-meson contributions are simply neglected in this work 
since two-body decay channels, which are open for any nucleon 
resonances, often play an important role for the dynamical properties 
of these resonances.
  We know, however, that multi-meson contributions have relatively large
effects to the several partial waves such as $P_{11}$.  We leave the 
calculations including the many-body contributions 
as a future subject. 

 Only $\pi$ and $\eta$ are considered for the flavor-octet mesons 
since nucleon resonances are dealt with in this paper.
As in Ref.\ \cite{GPP}, $\eta'$ is also taken into account.
 The observed values are used for the meson masses.
The flavor symmetry is now broken. It therefore should be understood 
that the coupling constant and the energy are properly taken for each 
component in Eq.\ (\ref{eqn:pv_coupling}),  although it is expressed in the 
flavor symmetric form.
The model then contains three meson-quark coupling constants, 
$g_{\pi qq}$,  $g_{\eta qq}$, and $g_{\eta ' qq}$.

 The mass operator for the single baryon $H_{\rm{eff}}(E)$ can be 
derived from the model Hamiltonian (\ref{eqn:hamiltonian}) by 
using the projection operator method \cite{ASY}.
 The contribution of meson-baryon continuum states is included in the 
energy-dependent effective potential $W(E)$.
 The mass operator is written as 
\begin{equation}
H_{\rm{eff}}(E) = T_{0} + V_{\rm{conf}} + W(E) + M_{0}\ ,
\label{eqn:massop}
\end{equation}
with
\begin{eqnarray}
\lefteqn{W(E) = H_{I} 
\frac{1}{E-H_{MB}+i\epsilon} 
H_{I}^{\dagger}}&&
\nonumber\\
&\equiv& H_{I} G(E) H_{I}^{\dagger}
\equiv \sum_{i,j} H_{I}(i) G(E) H_{I}^{\dagger}(j)\ ,
\end{eqnarray}
where $H_{MB} (=H-H_{I})$ is the total energy of  
intermediate meson-baryon states.

 Because the energy-dependent effective potential $W(E)$  
diverges if an infinite number of intermediate continuum states 
are rigorously taken into account, a prescription has to be 
introduced in order to perform the actual calculation \cite{Chin,Oset,HIN}.
 At first, the following operators are defined as
\begin{equation}
\bar{G} = -\frac{1}{\omega_{M}}\ ,
\end{equation}
\begin{eqnarray}
\bar{H}_{I}
&=&  
- \sum_{i=1}^3 \frac{1}{2m}\, \rho(\bfk)\, \bfsigma_i\cdot\bfk
\nonumber\\
&& \times
   \left( g_{Mqq}\,
     \bflambda_i\cdot\bfphi_i(\bfk)
     +\sqrt{\frac{2}{3}} g_{\eta'qq}\,
     \phi^{\eta'}_i(\bfk)
   \right)
 \nonumber\\
&\equiv& \sum_{i=1}^3 \bar{H}_{I}(i)\ .
\end{eqnarray}
 The operators $\bar{G}$ and $\bar{H}_{I}$ are obtained by 
applying the static approximation to $G(E)$ and $H_{I}$,
respectively. In this approximation, the baryons have the same
energy in the initial, final, and intermediate states.
 By using these auxiliary operators, the effective potential $W(E)$ 
can be decomposed into several parts as follows:
\begin{eqnarray}
W(E) &=& \sum_{i,j} H_{I}(i) G(E) H^{\dagger}_{I}(j)
   - \sum_{i,j} \bar{H}_{I}(i) \bar{G} \bar{H}^{\dagger}_{I}(j)
\nonumber\\
&&   + \sum_{i,j} \bar{H}_{I}(i) \bar{G} \bar{H}^{\dagger}_{I}(j)
\nonumber\\
&=& \sum_{i\neq j}\bar{H}_{I}(i)\bar{G}\bar{H}^{\dagger}_{I}(j)
   + \sum_{i,j} H_{I}(i) G(E) H^{\dagger}_{I}(j)
\nonumber\\
&&   - \sum_{i,j} \bar{H}_{I}(i) \bar{G} \bar{H}^{\dagger}_{I}(j)
   + \sum_{i=j} \bar{H}_{I}(i) \bar{G} \bar{H}^{\dagger}_{I}(j)
\nonumber\\
&\equiv& H_{\rm{OMEP}} + \Sigma(E) - \bar{\Sigma} + \bar{M}_{0}\ .
\label{eqn:dec_W}
\end{eqnarray}

 It can be easily shown that $H_{\rm {OMEP}}$ corresponds to the OMEP 
by the explicit calculation:
\begin{eqnarray}
\lefteqn{
H_{\rm{OMEP}}}\nonumber&&\\
 &=& \sum_{i<j}\left( \bftau_i\cdot\bftau_jV_{\pi}(r_{ij})
+\frac{1}{3}V_{\eta}(r_{ij})+\frac{2}{3} V_{\eta\prime}(r_{ij})\right)\ ,
\end{eqnarray}
with
\begin{eqnarray}
V_{M}(r_{ij}) &=& \frac{g_{Mqq}^2}{4\pi}\frac{1}{4m^2}\frac{1}{3}
\left[
S_{M}(r_{ij})\bfsigma_i\cdot\bfsigma_j
\right.
\nonumber\\
&&
\left.
  +T_{M}(r_{ij})
  \left(\frac{3(\bfsigma_i\cdot\bfr_{ij})
    (\bfsigma_j\cdot\bfr_{ij})}{r_{ij}^2}
    -\bfsigma_i\cdot\bfsigma_j
  \right)
\right]\ ,
\nonumber\\
\end{eqnarray}
where $r_{ij}$ is the relative separation between the $i$th and  
$j$th quarks, and $\bftau$ the isospin operator of quarks. 
 The space part of the spin-spin and tensor interactions of the OMEP
are denoted by $S_{M}$ and $T_{M}$, 
respectively. 
 They are explicitly written as follows: 
\begin{eqnarray}
S_{M}(r_{ij}) &=& 
\frac{2m_{M}^2}{\pi}\int_{0}^{\infty}dq
\frac{q^2}{q^2+m_{M}^2}\rho^2(q)j_{0}(qr_{ij})
\nonumber\\
&&-\frac{4}{\sqrt{\pi}} a^3e^{-a^2r_{ij}^2}\ ,
\label{spin-spin}
\end{eqnarray}
\begin{equation}
T_{M}(r_{ij}) = \frac{2}{\pi}
\int_{0}^{\infty}dq\frac{q^4}{q^2+m_{M}^2}\rho^2(q)j_2(qr_{ij})\ .
\end{equation}
 The second term of $S_{M}$, which stems from the $\delta$-function 
and is properly normalized, provides short-range interactions 
(see Eqs.\ (ref{eqn:hamiltonian1}) and (\ref{eqn:hamiltonian2}) 
in Ref.\ \cite{GPP}).
The OMEP, a part of the static contributions of the effective potential 
$W(E)$, is free from divergence although it includes all the 
intermediate states. 
 By specifying the contribution of the OMEP in $W(E)$, it becomes 
easy to clarify the correspondence between the present model and 
the conventional static models including the OMEP only \cite{GPP}.

 The second and third terms in Eq.\ (\ref{eqn:dec_W}) are the 
self-energy $\Sigma(E)$ 
and the subtraction term $\bar{\Sigma}$, respectively. 
The subtraction term plays a crucial role to get rid of double
counting, which otherwise should become a serious problem.
 In order to avoid the divergence in the mass operator $H_{\rm{eff}}(E)$, 
the cut-off of the intermediate states in these terms are introduced. 
 In the present work, only $\pi N$ and $\eta N$ states are taken into 
account because they are expected to have the largest contributions
among the meson-baryon continuum states. 
 The energy-dependence of the self-energy is quite important to 
reproduce the dynamical properties of baryons \cite{ASY,HAM,HMA},
while there is no energy-dependent quantity 
in the conventional models that contain the OMEP or the OGEP only. 
 The $\eta'N$ state, whose threshold is near $2000\,\rm{MeV}$, is not 
included since the resulting self-energy does not have  
strong energy-dependence in the energy region that we are concerned with. 

 The matrix element of the self-energy $\Sigma^{\pi N}_{ij}(E)$, 
which comes from the $\pi N$ intermediate state,   
is explicitly written as follows: 
\begin{equation}
\Sigma_{ij}^{\pi N}(E) =
\int\frac{d^3k}{(2\pi)^3}
\frac{\langle N_i^*|H_{I}|\pi N;\bfk\rangle
\langle\pi N;\bfk|H_{I}^{\dagger}|N_j^*\rangle}
{E-\sqrt{m_{\pi}^2+k^2}-\sqrt{m_N^2+k^2}+i\epsilon}\ ,
\label{eqn:self-energy}
\end{equation}
 where $|\pi N;\bfk\rangle$ is the $\pi N$ scattering state
with the relative momentum $\bfk$. 
The spin-flavor dependence of the self-energy is the same as that of
the OMEP since it essentially comes from 
the vertex function $\langle N^*|H_{I}|\pi N\rangle$, 
i.e., the matrix element of the meson-quark coupling. 
It has been indicated that this spin-flavor structure is important 
to resolve some long-standing problems in baryon 
spectroscopy \cite{Glozman-Riska}.
  In the integrand of Eq.\ (\ref{eqn:self-energy}), the relativistic 
form of $H_{MB}$ is used to avoid the unphysical momentum-dependence 
that the non-relativistic form has in the virtual high-momentum region.
For the nucleon mass, which should be calculated by the present model,
the observed value is temporarily used.
This is important to obtain the correct energy-dependence of the
self-energy. 
 The `$0\hbar\omega$' harmonic-oscillator wave function is used 
for the nucleon in the intermediate states.  
It has been found by numerical calculations that $90\,\%$ of 
the three-quark component of the nucleon consists of this 
`$0\hbar\omega$' wave function (see Section 3).

 The state-independent divergent quantity $\bar{M}_{0}$ should 
not affect the results and it should be cancelled by 
some counter term. 
 It is therefore considered in this work that the finite contribution of 
$\bar{M}_{0}$ is included in the bare mass $M_{0}$, and this quantity 
is removed from the potential. 
 
 Finally the mass operator is written as
\begin{eqnarray}
H_{\rm{eff}}(E) &=& T_{0} + V_{\rm{conf}} + H_{\rm{OMEP}}
+ \Sigma^{\pi N}(E) - \bar{\Sigma}^{\pi N} 
\nonumber\\
&&+ \Sigma^{\eta N}(E) - \bar{\Sigma}^{\eta N} + M_{0}\ .
\label{eqn:massop2}
\end{eqnarray}
 It should be emphasized that the decomposition of $W(E)$ does not 
lead to a simple sum of the OMEP and the self-energy. 
 In order to deal with  meson-quark coupling consistently and 
to avoid double counting, it is necessary to include 
the subtraction term $\bar{\Sigma}$.

 Before closing this section, we present the method of calculating 
the baryon mass spectrum.
 The matrix elements of the mass operator $H_{\rm{eff}}(E)$ are 
systematically calculated with the basis functions,
i.e., the three-quark bound-state wave functions, which are the 
antisymmetrized products of the quark wave functions
that consist of the space, spin, flavor and color parts.
 The spin, flavor and color wave functions are easily constructed 
and have well-defined symmetries. 
 For the space part, the harmonic-oscillator wave functions are used
with the range parameter $\beta$.
 These functions are convenient because they have analytic form
and also because the c.m. motion can be easily removed by using  
Jacobi coordinates.  
 The antisymmetrization can be done by using Talmi-Moshinsky 
coefficients \cite{OS,TM,TM2}. 

 The basis functions have to be truncated in practical 
applications. In the present case, the truncation requires the optimum 
choice of the parameter $\beta$, which determines the extension 
of wave functions.  
 The most appropriate value is searched by minimizing the energy 
eigenvalues of the static part of $H_{\rm{eff}}(E)$, i.e., 
$T_{0}+V_{\rm{conf}}+H_{\rm{OMEP}}+M_{0}$.
The obtained value  depends on the model parameters,
especially on the strength of the confinement potential.
 For the parameters given in table \ref{tbl:parameter} in the next section, 
this variational method provides the value of $\beta = 3.7\,\rm{fm}^{-2}$.
 It also has been found that the basis functions upto the  
$8\hbar\omega$-shell of the harmonic-oscillator wave functions 
are enough to obtain the results with good accuracy and stability.

The energy-dependence and the imaginary part of the 
mass operator $H_{\rm{eff}}(E)$  prevent us from naively
interpreting its eigenvalues as resonance masses. 
 The resonances correspond to the S-channel poles of the propagator  
for meson-baryon scattering: 
\begin{equation}
\hat{G}\propto\frac{1}{E-H_{\rm{eff}}(E)}\ .
\end{equation}
Therefore the resonance mass $E_{R}$ can be approximately 
determined  as the solution of 
\begin{equation} 
{\rm Re}(E_{R}-H_{\rm{eff}}(E_{R}))=0\ ,
\label{eqn:mass}
\end{equation}
after the energy-dependent eigenvalues are obtained by the  
diagonalization of the mass operator $H_{\rm{eff}}(E)$.

\section{Results and discussions}

  The present model still has six parameters to be determined:
the strength of the  confinement potential $c$, 
the form factor parameter $a$, 
the bare mass of baryons $M_{0}$, and  
the $\pi$-, $\eta$-, and $\eta'$-quark coupling constants, 
$g_{\pi qq}$, $g_{\eta qq}$, and $g_{\eta' qq}$, 
respectively.
 The parameters except $g_{\pi qq}$ are  phenomenologically determined 
to reproduce the prominent feature of the mass spectrum of the spin 
$\frac{1}{2}$ nucleon resonances:
The separations among the negative- and positive-parity resonances are 
relatively small in comparison with the large mass difference 
between the ground state nucleon and the other resonances. 
 All the parameters thus determined are summarized in 
table \ref{tbl:parameter}. 

 The $\pi$-quark coupling constant $g_{\pi qq}$ is derived from the 
$\pi N$ coupling constants $G_{\pi NN}$ by using the spin-flavor SU(6) 
relation of the quark model.
 The standard value of $G_{\pi NN}$ cited in Ref.\ \cite{compilation} 
is used:  $G_{\pi NN} = 14.3$.
For the determination of the 
$\eta$- and $\eta'$-quark coupling constants, $g_{\eta qq}$ and 
$g_{\eta' qq}$, the observed values of the meson-nucleon coupling
constants are not used 
because of the badly broken flavor SU(3) relation for these mesons. 
For $g_{\eta qq}$, for example, the fitting process provides the 
small value of 3.52, which is different from the value of 4.59 
derived from the $\eta$-nucleon coupling constant.

 The excitation energies put the constraint on the determination 
of the confinement strength $c$.
Because Eq.\ (\ref{eqn:mass}) is a non-linear
equation of energy variable, the 
role of $M_{0}$ is more than just an additional constant.
The phenomenological parameter $a$, which may be related 
with the finite size of the meson and the constituent quark, 
is quite difficult to be determined from other sources of 
information.
In the fitting procedure, this parameter  has correlation  
with the meson-quark coupling strengths through the momentum
dependence of the form factor.

\begin{figure}
\special{epsfile=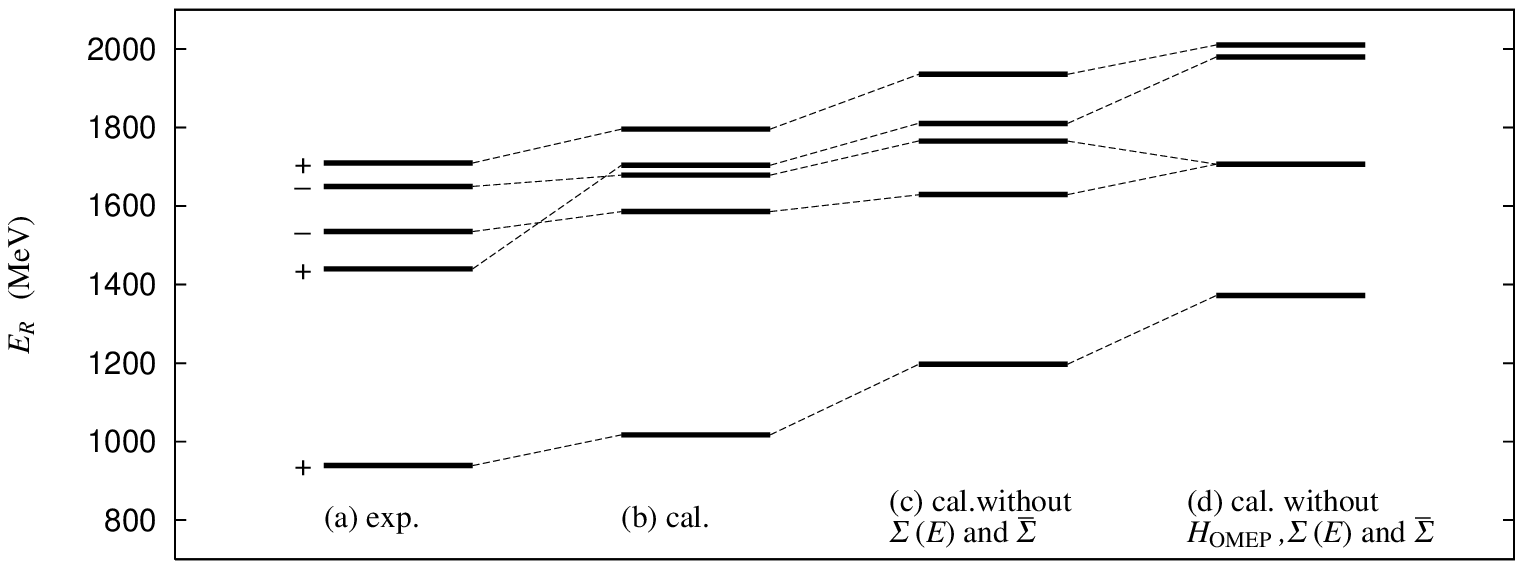}
\vspace{7cm}
\caption{
  The mass spectrum of nucleon resonances with spin $\frac{1}{2}$. 
  The calculations are compared with the experimental data. 
  The corresponding states are connected by dashed lines.
  The contributions of $H_{\rm{OMEP}}$, $\Sigma(E)$, and $\bar{\Sigma}$ 
are examined also.
 (a) The data taken from the particle data table \protect\cite{PDG}.
     The parities of the resonances are shown.
 (b) The result of the present model.
 (c) The result without  $\Sigma(E)$ and $\bar{\Sigma}$.
 (d) The result without  $H_{\rm{OMEP}}$, $\Sigma(E)$, and $\bar{\Sigma}$.
}
\label{fig1}
\end{figure}

 Let us present the mass spectra of the current model.
 The masses of the spin $\frac{1}{2}$ nucleon resonances are shown in 
Fig.\ \ref{fig1}. 
 The calculation  can reproduce the general feature of the 
observed spectrum fairly well.
 The excitation energies of the first negative-parity states 
are calculated to be about $500\,\rm{MeV}$, 
and the mass differences among the excited nucleons becomes 
relatively small compared with the excitation energies.

 In order to make clear the reason why the model can reproduce these 
features, each contribution of the mesonic effects on the mass 
spectrum is examined. As is seen in Fig.\ \ref{fig1}, where the result without 
the self-energy $\Sigma (E)$ and subtraction term  $\bar{\Sigma}$ 
is also shown, the self-energy, as well as  the OMEP, significantly 
contributes to the mass spectrum.
 In the consistent treatment of the meson-quark coupling in this work, 
the self-energy has almost the same magnitude as the OMEP. 
 For example, the diagonal element of the OMEP for the ground-state 
nucleon is $-175\,\rm{MeV}$, and the corresponding self-energy 
$\Sigma (E \simeq m_N)$ is about $-400\,\rm{MeV}$.  
  The subtraction term $\bar{\Sigma}$ is  
$-266\,\rm{MeV}$, and is comparable with the self-energy and the OMEP.
  Therefore the naive sum of the self-energy and the OMEP without
this subtraction causes serious overestimate of the mesonic effects.
It should be also emphasized that the subtraction term depends on the
initial and final baryon states and is not merely a constant parameter.

\begin{figure}
\special{epsfile=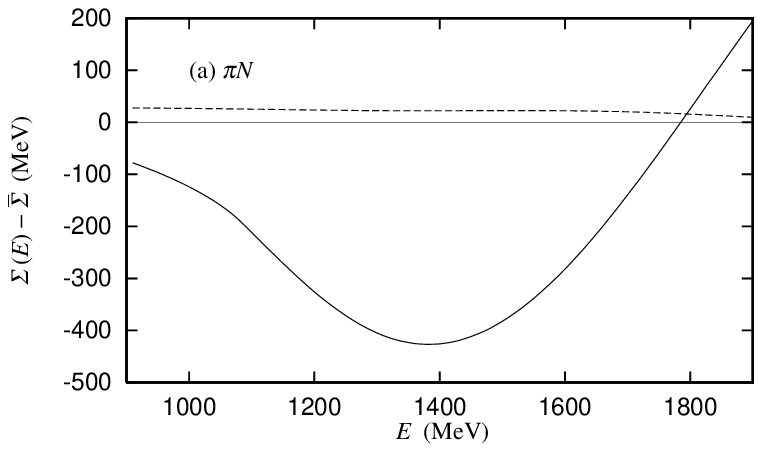}\hspace*{8.1cm}
\special{epsfile=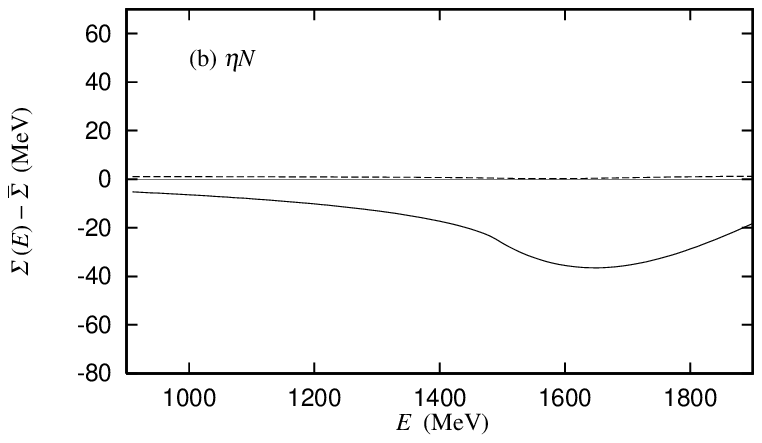}
\vspace{5cm}
\caption{
 (a) The matrix elements of 
$\Sigma^{\pi N}(E)-\bar{\Sigma}^{\pi N}$ 
for the positive-parity nucleon resonances with spin $\frac{1}{2}$.
 The solid line corresponds to the diagonal element of 
the $0\hbar\omega$ state, and the dashed line to that 
of the $2\hbar\omega$ state.
 Among the diagonal elements of the four $2\hbar\omega$ states, 
only the largest one is shown here.
 (b) Same as (a) but for 
$\Sigma^{\eta N}(E)-\bar{\Sigma}^{\eta N}$. 
}
\label{fig2}
\end{figure}

 For the positive-parity states, Fig.\ \ref{fig2} shows that
the self-energy due to the $\eta N$ state is generally smaller than 
that due to the $\pi N$ state. This is because of the restriction of the 
phase space since the $\eta N$ threshold is higher than the $\pi N$
threshold.
 The self-energy for the ground-state nucleon N is remarkably large 
because $N$ strongly couples to the intermediate $\pi N$ 
state without changing the configuration 
of quarks in a baryon.  The nucleon mass is pulled down partly 
owing to this effect.

\begin{figure}
\special{epsfile=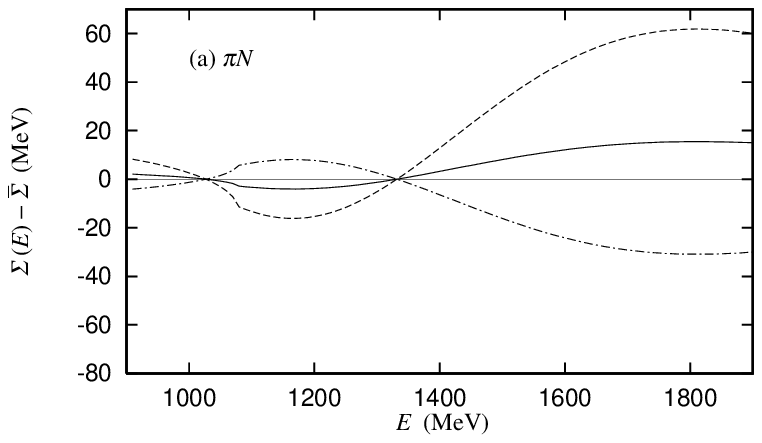}\hspace*{8.1cm}
\special{epsfile=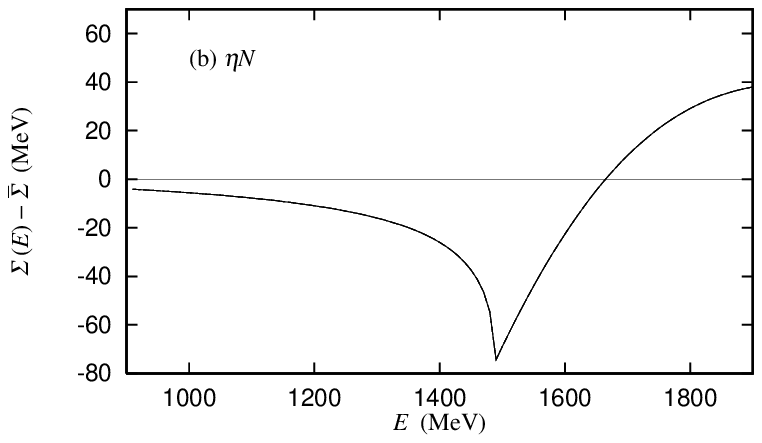}
\vspace{5cm}
\caption{
  Same as Fig.\ 2 but for the negative-parity nucleon resonances with 
spin $\frac{1}{2}$.
 The diagonal elements and the nondiagonal element for the 
two $1\hbar\omega$ states are shown. 
  The solid line corresponds to the diagonal element of the intrinsic-spin 
$\frac{3}{2}$ state, the dashed line to that of the 
intrinsic-spin $\frac{1}{2}$ state, 
and the dot-dashed line to the nondiagonal element between these states.
 Because the $\eta NN^*$ vertices for these negative-parity 
states are identical, all the matrix elements for the $\eta N$ coupling 
have the same value, that is, the three lines coincide with each other.
}
\label{fig3}
\end{figure}

In Fig.\ \ref{fig3}, the cusp behavior is  seen at the $\eta N$ threshold 
for the negative-parity states due to 
their $S$-wave coupling to the $\eta N$ state. 
 It is a characteristic feature of the self-energy, and this 
energy dependence is important to the dynamical properties of 
resonances.  The OMEP does not have such energy-dependence. 

  For the states in the same `$\hbar\omega$'-shell, the non-diagonal 
elements of the self-energy have the same magnitude as the diagonal elements.
  Note, however, that the state mixing among the different shells is 
relatively small since the off-diagonal elements of the total mass operator 
$H_{\rm{eff}}(E)$ are in general smaller than the differences in its diagonal
elements.
  
  In the present calculation, the mass of the ground-state nucleon $N$ 
is $1017\,\rm{MeV}$, which is larger than the observed one by about $10\,\%$.
 The disagreement is partly ascribed to the approximate treatment 
of $N$ in the intermediate $\pi N$ and $\eta N$ states.  In these decay 
channels, $N$ is considered just a three-quark bound state since multi-meson
states are neglected in this work. The physical nucleon is, however,
the admixture of the three-quark state, meson and three-quark state, 
and so on.
  The probability of observing the three-quark component   
is related with the energy-derivative of $\Sigma(E)$ as
\begin{equation}
Z^{-1} = 1-\rm{Re}\,\left(
\frac{d}{d E}\Sigma^{\pi N}(E)+
\frac{d}{d E}\Sigma^{\eta N}(E)
\right)_{E=E_{R}}\ .
\end{equation}
  Due to the strong energy-dependence of $\Sigma(E)$ around $E_{R}$,
the probability $Z$ deviates from 1: We obtain $Z \simeq 0.6$ for $N$.
  Because the relatively large part of the wave function is occupied by the 
meson-nucleon component, further study of the mesonic effects is 
required if we try to describe the property of the nucleon in detail.

  We examine the role of the spin-flavor symmetry of the OMEP by comparing 
it with the ``color-spin symmetry'' of the OGEP.
  To show the difference between the OMEP and the OGEP in an explicit way, 
we consider the $\frac{1}{2}^+$ and $\frac{1}{2}^-$ excited-states,
which are $2\hbar\omega$ and $1\hbar\omega$ excited-states, respectively.
  The three-quark bound-state with  
[3]$_{\rm space}$ $\otimes$ [21]$_{\rm spin}$ $\otimes$ 
[21]$_{\rm flavor}$ $\otimes$ [111]$_{\rm color}$  symmetry is taken 
for the $\frac{1}{2}^+$ state, and that with 
[21]$_{\rm space}$ $\otimes$ [21]$_{\rm spin}$ $\otimes$ 
[21]$_{\rm flavor}$ $\otimes$ [111]$_{\rm color}$ symmetry is taken  
for the $\frac{1}{2}^-$ state.
 Note that $N(1440)$ is dominated by this $\frac{1}{2}^+$ state.
 On the other hand, the symmetry property of the OMEP 
is characterized by 
$f\bflambda_i\cdot\bflambda_j\bfsigma_i\cdot\bfsigma_j$, 
where $f$ stands for the spatial part of the interaction, 
and $i$ and $j$ denote the $i\,$th and $j\,$th quarks, respectively.
Similarly, the OGEP has the operator $-g\,\bfsigma_i\cdot\bfsigma_j$.
 We obtain the diagonal matrix elements of these operators as 
shown in table \ref{tbl:matrix-element}.
  Since these potentials are of short-range and attractive, the following 
relations hold: 
$\langle f\rangle_{00} < \langle f\rangle_{10} < \langle f\rangle_{01} < 0$ 
and 
$\langle g\rangle_{00} < \langle g\rangle_{10} < \langle g\rangle_{01} < 0$.
 In the case of the OMEP, the difference is
$\frac{5}{2}\langle f \rangle_{10} + \frac{3}{5}\langle f \rangle_{01}$.
 Since the two terms are added constructively, the mass of  
the $\frac{1}{2}^+$ state is lowered more largely  
than that of the $\frac{1}{2}^-$ state due to the OMEP,
that is, the former gets close to the latter.  
In the case of the OGEP, on the other hand, the difference is 
$\frac{1}{2}(\langle g \rangle_{10} - \langle g \rangle_{01})$,
the magnitude of which becomes small owing to the destructive combination.
The similar argument can be applied to the other $\frac{1}{2}^-$ state 
with  [21]$_{\rm space}$ $\otimes$ [3]$_{\rm spin}$ $\otimes$ 
[21]$_{\rm flavor}$ $\otimes$ [111]$_{\rm color}$ symmetry. 
 The property of the spin-flavor symmetry in the OMEP is thus  
consistent with the fact the observed mass of $N(1440)$ is remarkably light.

  The spin-flavor symmetry of the self-energy also plays an important 
role in making the $\frac{1}{2}^+$ resonances come close 
to the $\frac{1}{2}^-$ resonances, as is shown in Fig.\ \ref{fig1}.
  The symmetry structure of the meson-quark coupling is characterized by the 
operator $\bfsigma_i \, \bflambda_i$ 
(see Eq.\ (\ref{eqn:pv_coupling})).
  Taking the same states (see table \ref{tbl:matrix-element}) for examples,  
we obtain the ratio of the spin-flavor parts of the $\pi NN^*$ vertices as
\begin{equation}
\frac{5\sqrt{2}}{4} 
= \frac{{\frac{1}{2}}^+\to\pi N {\rm transition}}
{{\frac{1}{2}}^- \to \pi N {\rm transition}}\ .
\end{equation}  
The ratio for the other $\frac{1}{2}^-$ state with [21]$_{\rm space}$ 
$\otimes$ [3]$_{\rm spin}$  $\otimes$ 
[21]$_{\rm flavor}$ $\otimes$ [111]$_{\rm color}$ is $5\sqrt{2}/2$ 
for spin-nonflip transition.
We can therefore expect that the mass shift of $N(1440)$ is larger than 
that of $N(1535)$ (see Eq.\ (\ref{eqn:self-energy})).
  More quantitative discussion needs proper treatment of the space part in 
each matrix element of the self-energy.
  The numerical calculations show that these mass shifts are also attractive 
and the spin-flavor symmetry of the self-energy is important to reproduce 
the nucleon mass spectrum.

  In spite of the favorable property of the spin-flavor symmetry in the
present model, $N(1440)$ still locates 
above the negative-parity resonances although it is 
experimentally observed below them.
  Glozman {\it et al.} have recently claimed that 
their OMEP model reproduces the mass of $N(1440)$ if
the relativistic kinematics is used for quarks \cite{GPVW}. (Although 
they can explain the mass spectrum with the non-relativistic 
kinematics in their first paper \cite{GPP}, the $\delta$-function part of 
the OMEP, i.e., the second term of Eq.\ (\ref{spin-spin}), 
has an artificially large strength.) 
If the relativistic correction in the kinetic energy operator 
is so important, the vertex functions should be also modified
in a consistent manner. We leave such a calculation with 
relativistic corrections as a future project. 

\begin{figure}
\special{epsfile=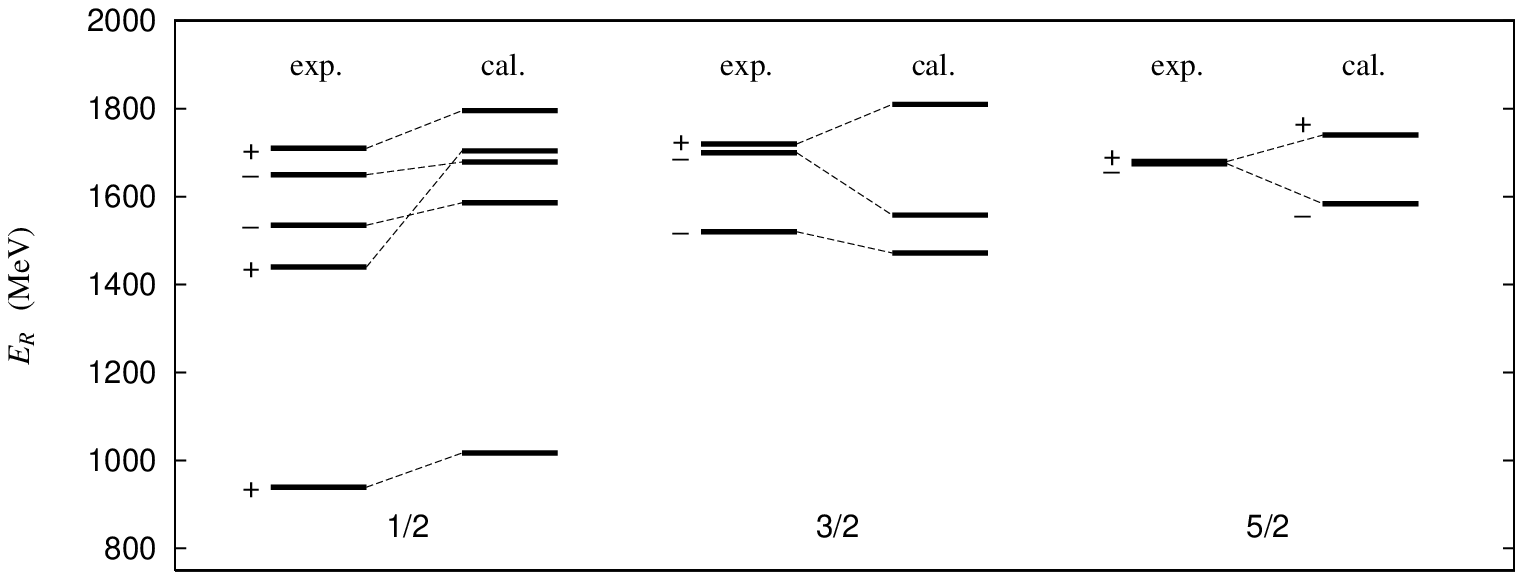}
\vspace{7cm}
\caption{
 The numerical results are compared with the observed spectra of nucleon 
resonances with spin $\frac{1}{2}$ (left), $\frac{3}{2}$ (center), and 
$\frac{5}{2}$ (right).
 The dashed lines represent the correspondence between the 
calculations and the data.
 The parities of the states are also shown.
}
\label{fig4}
\end{figure}

  We proceed to show the spectra of the spin $\frac{3}{2}$ and   
$\frac{5}{2}$ nucleon resonances in Fig.\ \ref{fig4}.
  All the parameters for these resonances are the same as those for 
the spin $\frac{1}{2}$ resonances. 
  The calculated masses of these resonances roughly agree 
with the observed values although the mass differences between the 
positive- and negative-parity states are too large.
  It should be noticed that the common bare mass $M_{0}$ can be used 
for all the nucleon resonances. 
  The form factor $\rho(\bfk)$ plays an important role in this success. 
  For example, the lowest $\frac{1}{2}^-$ and $\frac{3}{2}^-$ resonances, 
i.e., $N(1535)$ and $N(1520)$, are considered. 
  If $\rho(\bfk)$ is removed from the calculation of the self-energy, 
the mass of $N(1520)$ becomes smaller by about $60\,\rm{MeV}$, 
while that of $N(1535)$ changes by a few MeV only.
  This is because the $D$-wave $\pi NN(1520)$ coupling has large 
high-momentum component than the $S$-wave $\pi NN(1535)$ coupling.
The effect of $\rho(\bfk)$ is favorable since it reduces the contribution
from the virtual high-energy intermediate states that the present 
model may not be applied to.  
 But for $\rho(\bfk)$, the spin-dependence of $M_{0}$ should be 
inevitable.

\section{Summary}

 We have constructed the constituent quark model which contains the 
meson-quark coupling to incorporate the spontaneous breaking of 
chiral symmetry in low-energy QCD, and 
calculated the masses of nucleon resonances.
 The meson-quark coupling has the spin-flavor dependence which is 
thought to be important for the low-energy baryon spectrum.
 In the present formalism, the consistent treatment of the meson-quark 
coupling provides the OMEP and the self-energy of baryons with the 
subtraction term, which is important to avoid the problem of double counting. 
 Since both of the mesonic effects are significantly large,
we should take account of these terms simultaneously 
if we try to include the mesonic effects in the model of baryons.
 The results show that the model reproduces the gross feature of the 
observed mass spectra, whereas the mass of $N(1440)$ is 
still too high.
 
 In order to refine the calculations, we have to include not only
the $\pi N$ and $\eta N$ states but also the states that contain
other mesons and baryons, such as $\pi\Delta$ and $\rho N$.
These states, which are closely related to the $\pi\pi N$ channel, 
are considered to be important for the detailed study of 
nucleon resonance spectroscopy.  Relativistic corrections are 
to be investigated in near future also.

 Furthermore, we have to calculate the scattering quantities and 
compare them directly with the experimental data in order to 
complete the investigation of the model.
This type of study is necessary since the masses of nucleon 
resonances represent only a small part of the information of 
the phase shift analysis of the $\pi N$ scattering.
 The calculation of the scattering quantities are now in progress.
On the other hand, it is also desirable to study strange baryons 
in this model, and the results will be shown elsewhere.

\section*{Acknowledgements}
 We would like to thank Prof.\ R. Seki for valuable suggestions and
discussions. We are much indebted to Prof.\ T. Sato and Dr.\ T. Ogaito 
for discussions on the method of calculation.
 One of the authors (H.K.) would like to thank Prof.\ Y. Kudo and 
Prof.\ Y. Sakuragi for their continuous encouragement.

\newpage
\begin{table}[h]
\caption{
The values of the parameters.
The other parameters used in the calculations are as follows: 
$m = 340\,\rm{MeV}$, $\beta=3.7\,\rm{fm}^{-2}$, and
the observed values are used for the meson masses. 
}
\label{tbl:parameter}

\begin{center}
\begin{tabular}{cccccc}
\hline
\hline
$c\,(\rm{fm}^{-2})$ & $a\,(\rm{fm}^{-1})$ 
& $g_{\pi qq}$ & $g_{\eta qq}$ &
$g_{\eta' qq}$ & $M_{0}\,(\rm{MeV})$ \\
\hline
$1.5$ & $2.5$ & $2.91$ (fixed) & $3.52$ & $3.23$ & $0$ \\
\hline
\hline
\end{tabular}
\end{center}
\end{table}

\begin{table}[h]
\caption{
The matrix elements of the OMEP and the OGEP for 
$\frac{1}{2}^+$ and $\frac{1}{2}^-$ resonances.
The spatial matrix elements for the state with 
the node $n$ and the angular momentum $l$ are 
denoted by
$\langle f \rangle_{nl}$ and $\langle g \rangle_{nl}$.
}
\label{tbl:matrix-element}

\begin{center}
\begin{tabular}{ccc}
\hline
\hline
$J^{\pi}$ & OMEP & OGEP \\
\hline
$\frac{1}{2}^+$ & $\frac{5}{2}( \langle f \rangle_{00} +
\langle f \rangle_{10} )$ &
$\frac{1}{2}( \langle g \rangle_{00} +
\langle g \rangle_{10} )$ \\
\hline
$\frac{1}{2}^-$ & $\frac{5}{2}( \langle f \rangle_{00} -
\frac{3}{5}\langle f \rangle_{01} )$ &
$\frac{1}{2}( \langle g \rangle_{00} +
\langle g \rangle_{01} )$
\\
\hline
\hline
\end{tabular}
\end{center}
\end{table}

\newpage

\end{document}